# Haar Wavelet Based Approach for Image Compression and Quality Assessment of Compressed Image

Kamrul Hasan Talukder[I] and Koichi Harada[II]

*Abstract*— **With the increasing growth of technology and the entrance into the digital age, we have to handle a vast amount of information every time which often presents difficulties. So, the digital information must be stored and retrieved in an efficient and effective manner, in order for it to be put to practical use. Wavelets provide a mathematical way of encoding information in such a way that it is layered according to level of detail. This layering facilitates approximations at various intermediate stages. These approximations can be stored using a lot less space than the original data. Here a low complex 2D image compression method using wavelets as the basis functions and the approach to measure the quality of the compressed image are presented. The particular wavelet chosen and used here is the simplest wavelet form namely the Haar Wavelet. The 2D discrete wavelet transform (DWT) has been applied and the detail matrices from the information matrix of the image have been estimated. The reconstructed image is synthesized using the estimated detail matrices and information matrix provided by the Wavelet transform. The quality of the compressed images has been evaluated using some factors like Compression Ratio (CR), Peak Signal to Noise Ratio (PSNR), Mean Opinion Score (MOS), Picture Quality Scale (PQS) etc.**

## I. INTRODUCTION

The computer is becoming more and more powerful day by day. As a result, the use of digital images is increasing rapidly. Along with this increasing use of digital images comes the serious issue of storing and transferring the huge volume of data representing the images because the uncompressed multimedia (graphics, audio and video) data requires considerable storage capacity and transmission bandwidth. Though there is a rapid progress in mass storage density, speed of the processor and the performance of the digital communication systems, the demand for data storage capacity and data transmission bandwidth continues to exceed the capabilities of on hand technologies. Besides, the latest growth of data intensive multimedia based web applications has put much pressure on the researchers to find the way of using the images in the web applications more effectively. Internet teleconferencing, High Definition Television (HDTV), satellite communications and digital storage of movies are not feasible without a high degree of compression. As it is, such applications are far from realizing their full potential largely due to the limitations of common image compression techniques [1].

The image is actually a kind of redundant data i.e. it contains the same information from certain perspective of view. By using data compression techniques, it is possible to remove some of the redundant information contained in images. Image compression minimizes the size in bytes of a graphics file without degrading the quality of the image to an unacceptable level. The reduction in file size allows more images to be stored in a certain amount of disk or memory space. It also reduces the time necessary for images to be sent over the Internet or downloaded from web pages.

The scheme of image compression is not new at all. The discovery of Discrete Cosine Transform (DCT) in 1974 [2] is really an important achievement for those who work on image compression. The DCT can be regarded as a discrete time version of the Fourier Cosine series. It is a close relative of Discrete Fourier Transform (DFT), a technique for converting a signal into elementary frequency components. Thus DCT can be computed with a Fast Fourier Transform (FFT) like algorithm of complexity $O(n \log_2 n)$. Unlike DFT, DCT is real-valued and provides a better approximation of a signal with fewer coefficients.

There are a number of various methods in which image files can be compressed. There are two main common compressed graphic image formats namely Joint Photographic Experts Group (JPEG, usually pronounced as JAY-pehg) [3] and Graphic Interchange Format (GIF) for the use in the Internet. The JPEG method established by ISO (International Standards Organization) and IEC (International Electro-Technical Commission) is more often used for photographs, while the GIF method is commonly used for line art and other images in which geometric shapes are relatively simple.

In 1992, JPEG established the first international standard for still image compression where the encoders and decoders are DCT-based. The JPEG standard specifies three modes namely sequential, progressive, and hierarchical for lossy encoding, and one mode of lossless encoding. The performance of the coders for JPEG usually degrades at low bit-rates mainly because of the underlying block-based Discrete Cosine Transform (DCT) [4]. The baseline JPEG coder [5] is the sequential encoding in its simplest form. Fig. 1 and 2 show the key processing steps in such an encoder and decoder respectively for grayscale images. Color image compression can be approximately regarded as compression of multiple grayscale images, which are either compressed entirely one at a time, or are compressed by alternately interleaving 8x8 sample blocks from each in turn.

The DCT-based encoder can be thought of as essentially compression of a stream of 8x8 blocks of image samples. Each 8x8 block makes its way through each processing step, and yields output in compressed form into the data stream. Because adjacent image pixels are highly correlated, the Forward DCT (FDCT) processing step lays the basis for

[I]Currently a Phd student in the Department of Information Engineering of the Graduate School of Engineering, Hiroshima University, Japan (E-mail: khtalukder@hiroshima-u.ac.jp).

[II]Professor in the Graduate School of Engineering, Hiroshima University, Japan (E-mail: hrd@hiroshima-u.ac.jp).

gaining data compression by concentrating most of the signal in the lower spatial frequencies. For a typical 8x8 sample block from a typical source image, most of the spatial frequencies have zero or near-zero amplitude and need not to be encoded. Generally, the DCT introduces no loss to the source image samples; it merely transforms them to a domain in which they can be more efficiently encoded.

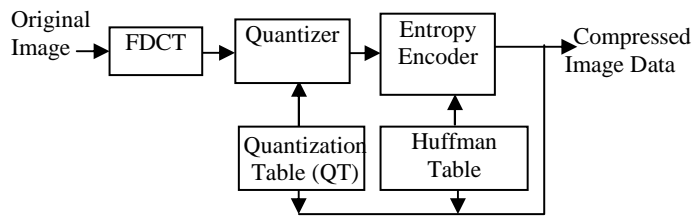

Fig.1: Encoder Block Diagram

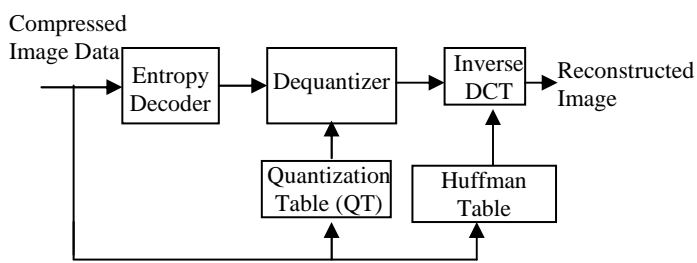

Fig. 2: Decoder Block Diagram

After output from the Forward DCT (FDCT), each of the 64 DCT coefficients is uniformly quantized in conjunction with a carefully designed 64-element Quantization Table (QT). At the decoder, the quantized values are multiplied by the corresponding QT elements to pick up the original unquantized values. After quantization, all the quantized coefficients are ordered into zig-zag sequence. This ordering helps to facilitate entropy encoding by placing low frequency non-zero coefficients before high-frequency coefficients. The DC coefficient, which contains a significant fraction of the total image energy, is differentially encoded.

Entropy Coding (EC) achieves additional compression losslessly through encoding the quantized DCT coefficients more compactly based on their statistical characteristics. The JPEG proposal specifies both Huffman coding and arithmetic coding.

More recently, the wavelet transform has emerged as a cutting edge technology, within the field of image analysis. Wavelets are a mathematical tool for hierarchically decomposing functions. Though rooted in approximation theory, signal processing, and physics, wavelets have also recently been applied to many problems in Computer Graphics including image editing and compression, automatic level-of-detail control for editing and rendering curves and surfaces, surface reconstruction from contours and fast methods for solving simulation problems in 3D modeling, global illumination, and animation [6]. Wavelet-based coding [7] provides substantial improvements in picture quality at higher compression ratios. Over the past few years, a variety of powerful and sophisticated wavelet-based schemes for image compression have been developed and implemented. Because of the many advantages of wavelet based image compression as listed below, the top contenders in the JPEG-2000 standard [8] are all wavelet-based compression algorithms.

- Wavelet coding schemes at higher compression avoid blocking artifacts.
- They are better matched to the HVS (Human Visual System) characteristics.
- Compression with wavelets is scalable as the transform process can be applied to an image as many times as wanted and hence very high compression ratios can be achieved.
- Wavelet based compression allow parametric gain control for image softening and sharpening.
- Wavelet-based coding is more robust under transmission and decoding errors, and also facilitates progressive transmission of images.
- Wavelet compression is very efficient at low bit rates.
- Wavelts provide an efficient decomposition of signals prior to compression.

II. BACKGROUND

Before we go into details of the method, we present some background topics of image compression which include the principles of image compression, the classification of compression methods and the framework of a general image coder and wavelets for image compression.

*A. Principles of Image Compression*

An ordinary characteristic of most images is that the neighboring pixels are correlated and therefore hold redundant information. The foremost task then is to find out less correlated representation of the image. Two elementary components of compression are redundancy and irrelevancy reduction. Redundancy reduction aims at removing duplication from the signal source image. Irrelevancy reduction omits parts of the signal that is not noticed by the signal receiver, namely the Human Visual System (HVS). In general, three types of redundancy can be identified: (a) Spatial Redundancy or correlation between neighboring pixel values, (b) Spectral Redundancy or correlation between different color planes or spectral bands and (c) Temporal Redundancy or correlation between adjacent frames in a sequence of images especially in video applications. Image compression research aims at reducing the number of bits needed to represent an image by removing the spatial and spectral redundancies as much as possible.

*B. Classification of Compression Technique*

There are two ways that we can consider for classifying compression techniques-lossless vs. lossy compression and predictive vs. transform coding.

**Lossless vs. Lossy compression:** In lossless compression schemes, the reconstructed image, after compression, is numerically identical to the original image. However lossless compression can only achieve a modest amount of compression. An image reconstructed following lossy compression contains degradation relative to the original. Often this is because the compression scheme completely discards redundant information. However, lossy schemes are capable of achieving much higher compression. Under normal viewing conditions, no visible loss is perceived (visually lossless).

**Predictive vs. Transform coding:** In predictive coding, information already sent or available is used to predict future values, and the difference is coded. Since this is done in the image or spatial domain, it is relatively simple to implement and is readily adapted to local image characteristics. Differential Pulse Code Modulation (DPCM) is one particular example of predictive coding. Transform coding, on the other hand, first transforms the image from its spatial domain representation to a different type of representation using some well-known transform and then codes the transformed values (coefficients). This method provides greater data compression compared to predictive methods, although at the expense of greater computation.

*C. Framework of General Image Compression Method*
A typical lossy image compression system is shown in Fig. 3. It consists of three closely connected components namely (a) Source Encoder, (b) Quantizer and (c) Entropy Encoder. Compression is achieved by applying a linear transform in order to decorrelate the image data, quantizing the resulting transform coefficients and entropy coding the quantized values.

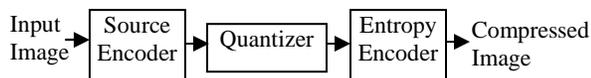

Fig. 3: A Typical Lossy Image Encoder

*Source Encoder (Linear Transformer)*
A variety of linear transforms have been developed which include Discrete Fourier Transform (DFT), Discrete Cosine Transform (DCT), Discrete Wavelet Transform (DWT) and many more, each with its own advantages and disadvantages.

*Quantizer*
A quantizer is used to reduce the number of bits needed to store the transformed coefficients by reducing the precision of those values. As it is a many-to-one mapping, it is a lossy process and is the main source of compression in an encoder. Quantization can be performed on each individual coefficient, which is called Scalar Quantization (SQ). Quantization can also be applied on a group of coefficients together known as Vector Quantization (VQ) [9]. Both uniform and non-uniform quantizers can be used depending on the problems.

*Entropy Encoder*
An entropy encoder supplementary compresses the quantized values losslessly to provide a better overall compression. It uses a model to perfectly determine the probabilities for each quantized value and produces an appropriate code based on these probabilities so that the resultant output code stream is smaller than the input stream. The most commonly used entropy encoders are the Huffman encoder and the arithmetic encoder, although for applications requiring fast execution, simple Run Length Encoding (RLE) is very effective [10].

It is important to note that a properly designed quantizer and entropy encoder are absolutely necessary along with optimum signal transformation to get the best possible compression.

*D. Wavelets for image compression*
Wavelet transform exploits both the spatial and frequency correlation of data by dilations (or contractions) and translations of mother wavelet on the input data. It supports the multiresolution analysis of data i.e. it can be applied to different scales according to the details required, which allows progressive transmission and zooming of the image without the need of extra storage. Another encouraging feature of wavelet transform is its symmetric nature that is both the forward and the inverse transform has the same complexity, building fast compression and decompression routines. Its characteristics well suited for image compression include the ability to take into account of Human Visual System's (HVS) characteristics, very good energy compaction capabilities, robustness under transmission, high compression ratio etc.

The implementation of wavelet compression scheme is very similar to that of subband coding scheme: the signal is decomposed using filter banks. The output of the filter banks is down-sampled, quantized, and encoded. The decoder decodes the coded representation, up-samples and recomposes the signal.

Wavelet transform divides the information of an image into approximation and detail subsignals. The approximation subsignal shows the general trend of pixel values and other three detail subsignals show the vertical, horizontal and diagonal details or changes in the images. If these details are very small (threshold) then they can be set to zero without significantly changing the image. The greater the number of zeros the greater the compression ratio. If the energy retained (amount of information retained by an image after compression and decompression) is 100% then the compression is lossless as the image can be reconstructed exactly. This occurs when the threshold value is set to zero, meaning that the details have not been changed. If any value is changed then energy will be lost and thus lossy compression occurs. As more zeros are obtained, more energy is lost. Therefore, a balance between the two needs to be found out.

III. HAAR WAVELET TECHNIQUE

*A. Haar Wavelet Transform*
To understand how wavelets work, let us start with a simple example. Assume we have a 1D image with a resolution of four pixels, having values [9 7 3 5]. Haar wavelet basis can be used to represent this image by computing a wavelet transform. To do this, first the average the pixels together, pairwise, is calculated to get the new lower resolution image with pixel values [8 4]. Clearly, some information is lost in this averaging process. We need to store some *detail coefficients* to recover the original four pixel values from the two averaged values. In our example, 1 is chosen for the first detail coefficient, since the average computed is 1 less than 9 and 1 more than 7. This single number is used to recover the first two pixels of our original four-pixel image. Similarly, the second detail coefficient is -1, since 4 + (-1) = 3 and 4 - (-1) = 5. Thus, the original image is decomposed into a lower resolution (two-pixel) version and a pair of detail coefficients.

Repeating this process recursively on the averages gives the full decomposition shown in Table I:

Table I: Decomposition to lower resolution

| Resolution | Averages | Detail Coefficients |
|---|---|---|
| 4 | [9 7 3 5] | |
| 2 | [8 4] | [1 -1] |
| 1 | [6] | [2] |

Thus, for the one-dimensional Haar basis, the wavelet transform of the original four-pixel image is given by [6 2 1 -1]. We call the way used to compute the wavelet transform by recursively averaging and differencing coefficients, *filter bank*. We can reconstruct the image to any resolution by recursively adding and subtracting the detail coefficients from the lower resolution versions.

*B. Compression of 2D image with Haar Wavelet Technique*

It has been shown in previous section how one dimensional image can be treated as sequences of coefficients. Alternatively, we can think of images as piecewise constant functions on the half-open interval [0, 1). To do so, the concept of a *vector space* is used. A one-pixel image is just a function that is constant over the entire interval [0, 1). Let $V^0$ be the vector space of all these functions. A two pixel image has two constant pieces over the intervals [0, 1/2) and [1/2, 1). We call the space containing all these functions $V^1$. If we continue in this manner, the space $V^j$ will include all piecewise-constant functions defined on the interval [0, 1) with constant pieces over each of $2^j$ equal subintervals. We can now think of every one-dimensional image with $2^j$ pixels as an element, or vector, in $V^j$. Note that because these vectors are all functions defined on the unit interval, every vector in $V^j$ is also contained in $V^{j+1}$. For example, we can always describe a piecewise constant function with two intervals as a piecewise-constant function with four intervals, with each interval in the first function corresponding to a pair of intervals in the second. Thus, the spaces $V^j$ are nested; that is, $V^0 \subset V^1 \subset V^2 \subset \ldots$. This nested set of spaces $V^j$ is a necessary ingredient for the mathematical theory of multiresolution analysis [6]. It guarantees that every member of $V^0$ can be represented exactly as a member of higher resolution space $V^1$. The converse, however, is not true: not every function $G(x)$ in $V^1$ can be represented exactly in lower resolution space $V^0$; in general there is some lost detail [11].

Now we define a basis for each vector space $V^j$. The basis functions for the spaces $V^j$ are called *scaling functions*, and are usually denoted by the symbol $\phi$. A simple basis for $V^j$ is given by the set of scaled and translated box functions [7]:

$$\phi_i^j(x) := \phi(2^j x - i) \quad i = 0, 1, 2 \ldots 2^j - 1 \text{ where}$$

$$\phi(x) := \begin{cases} 1 & \text{for } 0 \leq x < 1 \\ 0 & \text{otherwise} \end{cases}$$

The wavelets corresponding to the box basis are known as the *Haar wavelets*, given by-

$$\Psi_i^j(x) := \Psi(2^j x - i) \quad i = 0, 1, 2 \ldots 2^j - 1 \text{ where}$$

$$\Psi(x) := \begin{cases} 1 & \text{for } 0 \leq x < 1/2 \\ -1 & \text{for } 1/2 \leq x < 1 \\ 0 & \text{otherwise} \end{cases}$$

Thus, the DWT for an image as a 2D signal will be obtained from 1D DWT. We get the scaling function and wavelet function for 2D by multiplying two 1D functions. The scaling function is obtained by multiplying two 1D scaling functions: $\phi(x,y) = \phi(x)\phi(y)$. The wavelet functions are obtained by multiplying two wavelet functions or wavelet and scaling function for 1D. For the 2D case, there exist three wavelet functions that scan details in horizontal $\Psi^{(1)}(x,y) = \phi(x)\Psi(y)$, vertical $\Psi^{(2)}(x,y) = \Psi(x)\phi(y)$ and diagonal directions: $\Psi^{(3)}(x,y) = \Psi(x)\Psi(y)$. This may be represented as a four channel perfect reconstruction filter bank as shown in Fig. 4. Now, each filter is 2D with the subscript indicating the type of filter (HPF or LPF) for separable horizontal and vertical components. By using these filters in one stage, an image is decomposed into four bands. There exist three types of detail images for each resolution: horizontal (HL), vertical (LH), and diagonal (HH). The operations can be repeated on the low low (LL) band using the second stage of identical filter bank. Thus, a typical 2D DWT, used in image compression, generates the hierarchical structure shown in Fig. 5.

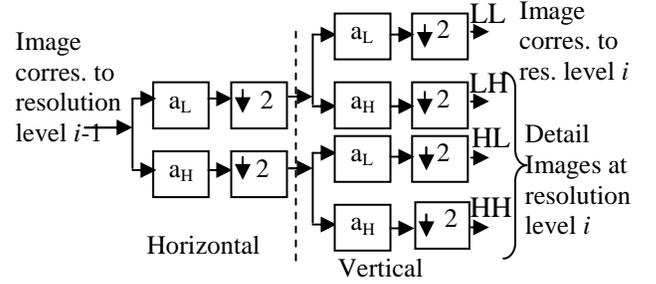

Fig. 4. One Filter Stage in 2D DWT

| LL | HL3 | | |
|---|---|---|---|
| LH3 | HH3 | HL2 | HL1 |
| LH2 | | HH2 | |
| LH1 | | | HH1 |

Fig. 5. Structure of wavelet decomposition

The transformation of the 2D image is a 2D generalization of the 1D wavelet transformed already discussed. It applies the 1D wavelet transform to each row of pixel values. This operation provides us an average value along with detail coefficients for each row. Next, these transformed rows are treated as if they were themselves an image and apply the 1D transform to each column. The resulting values are all detail coefficients except a single overall average co-efficient. In order to complete the transformation, this process is repeated recursively only on the quadrant containing averages.

Now let us see how the 2D Haar wavelet transformation is performed. The image is comprised of pixels represented by numbers [12]. Consider the 8×8 image taken from a specific portion of a typical image shown in Fig. 6. The matrix (a 2D array) representing this image is shown in Fig. 7.

Now we perform the operation of averaging and differencing to arrive at a new matrix representing the same image in a more concise manner. Let us look how the operation is done. Consider the first row of the Fig. 7.

Averaging: (64+2)/2=33, (3+61)/2=32, (60+6)/2=33, (7+57)/2=32

Differencing: 64–33 =31, 3–32= –29, 60–33=27 and 7–32= –25

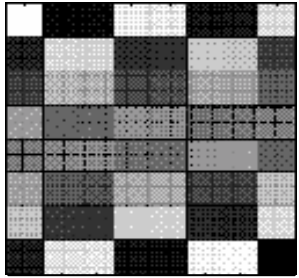

Fig. 6: A 8×8 image

$$\begin{pmatrix} 64 & 2 & 3 & 61 & 60 & 6 & 7 & 57 \\ 9 & 55 & 54 & 12 & 13 & 51 & 50 & 16 \\ 17 & 47 & 46 & 20 & 21 & 43 & 42 & 24 \\ 40 & 26 & 27 & 37 & 36 & 30 & 31 & 33 \\ 32 & 34 & 35 & 29 & 28 & 38 & 39 & 25 \\ 41 & 23 & 22 & 44 & 45 & 19 & 18 & 48 \\ 49 & 15 & 14 & 52 & 53 & 11 & 10 & 56 \\ 8 & 58 & 59 & 5 & 4 & 62 & 63 & 1 \end{pmatrix}$$

Fig. 7 : 2D array representing the Fig. 6

So, the transformed row becomes (33 32 33 32 31 –29 27 –25). Now the same operation on the average values i.e. (33 32 33 32) is performed. Then we perform the same operation on the averages i.e. first two elements of the new transformed row. Thus the final transformed row becomes (32.5 0 0.5 0.5 31 –29 27 –25). The new matrix we get after applying this operation on each row of the entire matrix of Fig. 7 is shown in Fig. 8. Performing the same operation on each column of the matrix in Fig. 8, we get the final transformed matrix as shown in Fig. 9. This operation on rows followed by columns of the matrix is performed recursively depending on the level of transformation meaning the more iteration provides more transformations. Note that the left-top element of the Fig. 9 i.e. 32.5 is the only averaging element which is the overall average of all elements of the original matrix and the rest all elements are the details coefficients. The main part of the C program used to transform the matrix is shown in Fig. 10. The 2D array *mat* holds the values which represent the image.

The point of the wavelet transform is that regions of little variation in the original image manifest themselves as small or zero elements in the wavelet transformed version. The 0's in the Fig. 9 are due to the occurrences of identical adjacent elements in the original matrix. A matrix with a high proportion of zero entries is said to be *sparse*. For most of the image matrices, their corresponding wavelet transformed versions are much sparser than the originals. Very sparse matrices are easier to store and transmit than ordinary matrices of the same size. This is because the sparse matrices can be specified in the data file solely in terms of locations and values of their non-zero entries.

$$\begin{pmatrix} 32.5 & 0 & 0.5 & 0.5 & 31 & -29 & 27 & -25 \\ 32.5 & 0 & -0.5 & -0.5 & -23 & 21 & -19 & 17 \\ 32.5 & 0 & -0.5 & -0.5 & -15 & 13 & -11 & 9 \\ 32.5 & 0 & 0.5 & 0.5 & 7 & -5 & 3 & -1 \\ 32.5 & 0 & 0.5 & 0.5 & -1 & 3 & -5 & 7 \\ 32.5 & 0 & -0.5 & -0.5 & 9 & -11 & 13 & -15 \\ 32.5 & 0 & -0.5 & -0.5 & 17 & -19 & 21 & -23 \\ 32.5 & 0 & 0.5 & 0.5 & -25 & 27 & -29 & 31 \end{pmatrix}$$

Fig. 8.: Transformed array after operation on each row of Fig. 7

$$\begin{pmatrix} 32.5 & 0 & 0 & 0 & 0 & 0 & 0 & 0 \\ 0 & 0 & 0 & 0 & 0 & 0 & 0 & 0 \\ 0 & 0 & 0 & 0 & 4 & -4 & 4 & -4 \\ 0 & 0 & 0 & 0 & 4 & -4 & 4 & -4 \\ 0 & 0 & 0.5 & 0.5 & 27 & -25 & 23 & -21 \\ 0 & 0 & -0.5 & -0.5 & -11 & 9 & -7 & 5 \\ 0 & 0 & 0.5 & 0.5 & -5 & 7 & -9 & 11 \\ 0 & 0 & -0.5 & -0.5 & 21 & -23 & 25 & -27 \end{pmatrix}$$

Fig. 9.: Final Transformed Matrix after one step

It can be seen that in the final transformed matrix, we find a lot of entries zero. From this transformed matrix, the original matrix can be easily calculated just by the reverse operation of averaging and differencing i.e. the original image can be reconstructed from the transformed image without the loss of information. Thus, it yields a lossless compression of the image. However, to achieve more degree of compression, we have to think of the lossy compression. In this case, a nonnegative threshold value say $\varepsilon$ is set. Then any detail coefficient in the transformed data whose magnitude is less than or equal to $\varepsilon$ is set to zero. It will increase the number of 0's in the transformed matrix and thus the level of compression is increased. So, $\varepsilon =0$ is used for a lossless compression. If the lossy compression is used, the approximations of the original image can be built up. The setting of the threshold value is very important as there is a tradeoff between the value of $\varepsilon$ and the quality of the compressed image. The different thresholding methods we have used are: hard thresholding, soft thresholding and universal thresholding. These thresholding methods are defined as follows:

$$T(\varepsilon,x)=\begin{cases} 0, & \text{if } |x|<\varepsilon \\ x, & \text{otherwise} \end{cases} \ldots\ldots\ldots\ldots \text{(Hard Thresholding)}$$

$$T(\varepsilon,x)=\begin{cases} 0, & \text{if } |x|<\varepsilon \\ \text{Sign}(x)(|x|-\varepsilon), & \text{otherwise} \end{cases} \ldots\ldots \text{(Soft Thresholding)}$$

$$T(\varepsilon,x)=\begin{cases} 0, & \text{if } x<\sigma(2log_2 \text{ N})^{1/2} \\ x, & \text{otherwise} \end{cases} \ldots\ldots \text{(Universal Thresholding)}$$

where $\sigma$ is the standard deviation of the wavelet coefficients and $N$ is is the number of wavelet coefficients.

Loosely saying, the compression ratio of the image is calculated by- the number of nonzero elements in original matrix : the number of nonzero elements in updated transformed matrix [13].

In summary, the main steps of the 2D image compression using Haar Wavelet as the basis functions are: (a) Start with the matrix *P* representing the original image, (b) Compute the transformed matrix *T* by the operation averaging and differencing (First for each row, then for each column) (c) Choose a threshold method and apply that to find the new matrix say *D* (e) Use *D* to compute the compression ratio and others values and to reconstruct the original image as well.

Now we see the effect of one step averaging and differencing of an image. The Fig. 11 (a) is the original image and the Fig. 11 (b) is the transformed image after applying the one step averaging and differencing. The more steps produce more decomposition.

```
/*row transformation*/
for(i=0;i<row;i++){w=col;
        do{ k=0;
/*averaging*/    for(j=0;j<w/2;j++)
                a[j]=((mat[i][j+j]+mat[i][j+j+1])/2);
/*differencing*/ for(j=w/2;j<w;j++,k++)
                a[j]=mat[i][j-w/2+k]-a[k];
            for(j=0;j<row;j++) mat[i][j]=a[j];
            w=w/2;
        }while(w!=1);
}
/*column transformation*/
for(i=0;i<col;i++){ w=row;
        do{k=0;
/*averaging*/    for(j=0;j<w/2;j++)
                a[j]=((mat[j+j][i]+mat[j+j+1][i])/2);
/*differencing*/for(j=w/2;j<w;j++,k++)
                a[j]=mat[j-w/2+k][i]-a[k];
            for(j=0;j<w;j++) mat[j][i]=a[j];
            w=w/2;
        }while(w!=1);
}
```

Fig. 10.: Code for the transformation

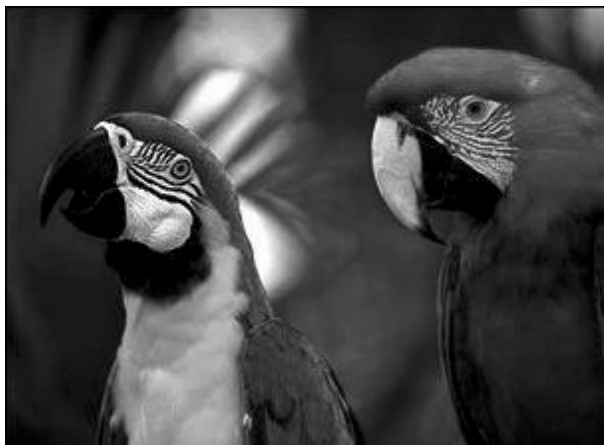

Fig. 11 (a): 1-level decomposition

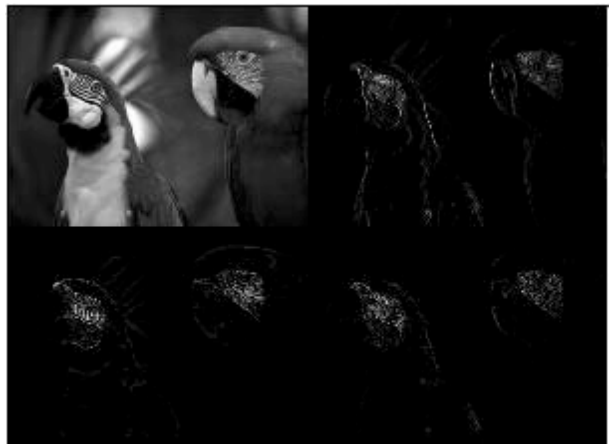

Fig. 11 (b): 2-level decomposition

IV. QUALITY MEASUREMENT

A database of twenty gray scale images each with size 256×256 is used in the experiment.

We define the *compression ratio* (CR) as the ratio of the number of nonzero elements in original matrix to the number of nonzero elements in updated transformed matrix. The enthusiastic CR values for different thresholding methods and different $\varepsilon$ is tabulated below in Table II:

Table II: Compression Ratio

| $\varepsilon$ | CR (Hard Threshold) | CR (Soft Threshold) |
|---|---|---|
| 15 | 15.74 | 14.1 |
| 20 | 17.11 | 15.87 |
| 25 | 18.47 | 16.95 |

The universal thresholding method generates the CR as 14.375. It is noted here that the hard thresholding provides the best CR. The soft thresholding gives better CR in comparison to universal thresholding method but it depends on choosing the value of $\varepsilon$.

The PSNR for gray scale image (8 bits/pixel) is defined by-

$$PSNR(dB) = 20 \times \log_{10}(\frac{255}{\sqrt{MSE}})$$

where MSE is the Man Squared Error defined by-

$$MSE = \frac{1}{mn}\sum_{y=1}^{m}\sum_{x=1}^{n}\left(I(x,y) - I^1(x,y)\right)^2$$

where $I$ is original image, $I^1$ is approximation of decompressed image and m, n are dimensions of the image. The PSNR values for different threshold values and techniques are shown in Fig. 12. The soft thresholding method performs better than hard thresholding. The universal method reports PSNR as 24.875. These results are very much acceptable in most cases except in medical application where no loss of information is to be guaranteed.

However, the PSNR is not adequate as a perceptually meaningful measure of picture quality, because the reconstruction errors generally do not have the characteristic of signal independent additive noise and the seriousness of the impairments cannot be measured by a simple power measurement. Small impairment of an image can lead to a very small PSNR in lieu of the fact that the perceived image quality can be acceptable. So, the perceptual quality measurement method quantified by MOS and PQS has been applied. The reference and test conditions are arranged in pairs such that the first is the unimpaired reference and the second is the same sequence impaired. The original image without com-pression was used as the reference condition. The viewers are asked to vote on the second, keeping in mind the first. The method uses the five grade impairment scale: 5 (Excellent), 4 (Good), 3 (Slightly annoying), 2 (Annoying) and 1 (Very annoying). At the end, the MOS is calculated as-

$$MOS = \sum_{i=1}^{5} i.p(i)$$

where $i$ is grade and $p(i)$ is grade probability. PQS defined by-

$$PQS = b_0 + \sum_{i=1}^{3} b_i Z_i$$

uses some properties of HVS relevant to global image impairments such as random errors and emphasizes the perceptual importance of structured and localized errors. Here, a linear combination of uncorrelated principal distortion measures $Z_i$ combined by partial regression coefficients $b_i$ are used. PQS is constructed by regressions with MOS. The MOS and PQS values obatined are tabulated below in Table III which are very much encouraging.

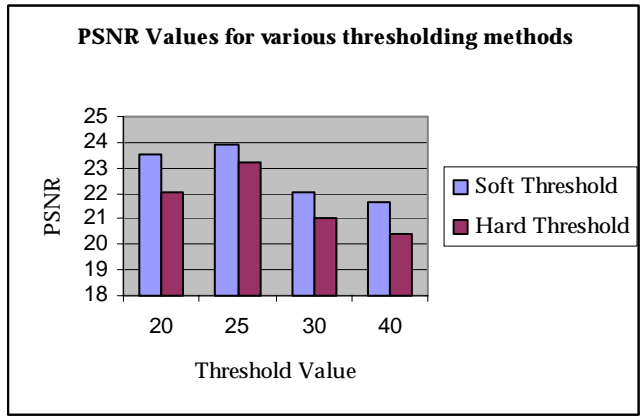

Fig. 12.: PSNR values for various thresholding

Table III: MOS and PQS values

| ε=15 | MOS | PQS |
|---|---|---|
| Hard Thresholding | 4.675 | 4.765 |
| Soft Thresholding | 4.80 | 4.875 |
| Universal Thresholding | 4.865 | 4.957 |

V. DISCUSSION

The number of decompositions determines the quality of compressed image. The number of decompositions also determines the resolution of the lowest level in wavelet domain. If a larger number of decompositions is used, it will provide more success in resolving important DWT coefficients from less important coefficients. The HVS is less sensitive to removal of smaller details. After decomposing the image and representing it with wavelet coefficients, compression can be performed by ignoring all coefficients below some threshold. In our experiment, compression is obtained by wavelet coefficient thresholding using different thresholding techniques like hard thresholding, soft thresolding and universal thresholding. All coefficients below some threshold are neglected and compression ratio is computed. Compression algorithm provides two modes of operation: 1) compression ratio is fixed to the required level and threshold value has been changed to achieve required compression ratio; after that, PSNR is computed; 2) PSNR is fixed to the required level and threshold values has been changed to achieve required PSNR; after that, CR is computed. It is noted that image quality is better for a larger number of decompositions. On the contrary, a larger number of decompositions causes the loss of the coding algorithm efficiency. Therefore, adaptive decomposition is required to achieve balance between image quality and computational complexity. PSNR tends to saturate for a larger number of decompositions. For each compression ratio, the PSNR characteristic has "threshold" which represents the optimal number of decompositions. Below and above the threshold, PSNR decreases.

At present, the most widely used objective distortion measures are the MSE and the related PSNR. They can easily be computed to represent the deviation of the distorted image from the original image in the pixelwise sense. However, in practical viewing situations, human beings are usually not concentrated on pixel differences alone, except for particular applications such as medical imaging, where pixelwise precision can be very important. The subjective perceptual quality includes surface smoothness, edge sharpness and continuity, proper background noise level, and so on. Image compression techniques induce various types of visual artifacts that affect the human viewing experience in many distinctways, even if the MSE or PSNR level is adjusted to be about equal. It is generally agreed that MSE or PSNR does not correlate well with the visual quality perceived by human beings, since MSE is computed by adding the squared differences of individual pixels without considering the spatial interaction among adjacent pixels. Some work tries to modify existing quantitative measures to accommodate the factor of human visual perception. One approach is to improve MSE by putting different weights to neighboring regions with different distances to the focal pixel [14]. Most approaches can be viewed as curve-fitting methods to comply with the rating scale method. In order to obtain an objective measure for perceived image fidelity, models of the human visual system (HVS) should be taken into account. It is well known that the HVS has different sensitivities to signals of different frequencies. Since the detection mechanisms of the HVS have localized responses in both the space and frequency domains, neither the space-based MSE nor the global Fourier analysis provides a good tool for the modeling. So, here the perceptual quality measurement method quantified by MOS and PQS has been applied and the results are encouraging.

The fundamental difficulty in testing an image compression system is how to decide which test images to use for evaluation. The image content being viewed influences the perception of quality irrespective of technical parameters of the compression system. A series of pictures which are average in terms of how difficult they are for system being evaluated, has been selected.

In this paper, only the gray-scale images are considered. However, wavelet transforms and compression techniques are equally applicable to color images with three color components. We have to perform the wavelet transform independently on each of the three color components of the images and have to treat the results as an array of vectored-valued wavelet co-efficients. In this case, in lieu of using the absolute value of the scalar co-efficient, a vector-valued co-efficient is to be used. Furthermore, a number of ways can be used in which the color information can be used to obtain a wavelet transform that is even sparser. For example, by first converting the pixel values in an image from RGB colors to YIQ colors [15], we can separate the luminance information ($Y$) from chromatic information ($I$ and $Q$). Once the wavelet transform is computed, the compression method can be applied to each of the components of the image separately. Since the human perception is most sensitive to variation in $Y$ and least sensitive in $Q$, the compression scheme may be permitted to tolerate a larger error in the $Q$ component of the compressed image, thereby increasing the scale of compression.

## VI. CONCLUSION

A picture can say more than a thousand words. However, storing an image can cost more than a million words. This is not always a problem because now computers are capable enough to handle large amounts of data. However, it is often desirable to use the limited resources more efficiently. For instance, digital cameras often have a totally unsatisfactory amount of memory and the internet can be very slow. In these cases, the importance of the compression of image is greatly felt. The rapid increase in the range and use of electronic imaging justifies attention for systematic design of an image compression system and for providing the image quality needed in different applications. Wavelet can be effectively used for this purpose. A low complex 2D image compression method using Haar wavelets as the basis functions along with the quality measurement of the compressed images have been presented here. As for the further work, the tradeoff between the value of the threshold $\varepsilon$ and the image quality can be studied and also fixing the correct threshold value is also of great interest. Furthermore, finding out the exact number of transformation level required in case of application specific image compression can be studied. Also, more thorough comparison of various still image quality measurement algorithms may be conducted. Though many published algorithms left a few parameters unspecified, here good estimates of them for implementation have been provided. All these metrics, including ours, did very well in estimating the perceptual error, so that it is difficult to conclude any decisive advantage of one algorithm over another.

REFERENCES

[1] Talukder, K.H. and Harada, K., *A Scheme of Wavelet Based Compression of 2D Image*, Proc. IMECS, Hong Kong, pp. 531-536, June 2006.
[2] Ahmed, N., Natarajan, T., and Rao, K. R., *Discrete Cosine Transform*, IEEE Trans. Computers, vol. C-23, Jan. 1974, pp. 90-93.
[3] Pennebaker, W. B. and Mitchell, J. L. JPEG, *Still Image Data Compression Standards*, Van Nostrand Reinhold, 1993.
[4] Rao, K. R. and Yip, P., *Discrete Cosine Transforms - Algorithms, Advantages, Applications*, Academic Press, 1990.
[5] Wallace, G. K., *The JPEG Still Picture Compression Standard*, Comm. ACM, vol. 34, no. 4, April 1991, pp. 30-44.
[6] Eric J. Stollnitz, Tony D. Derose and David H. Salesin, *Wavelets for Computer Graphics- Theory and Applications Book*, Morgan Kaufmann Publishers, Inc. San Francisco, California.
[7] Vetterli, M. and Kovacevic, J., *Wavelets and Subband Coding*, Englewood Cliffs, NJ, Prentice Hall, 1995, http://cm.bell-labs.com/who/jelena/Book/home.html.
[8] ISO/IEC/JTC1/SC29/WG1 N390R, *JPEG 2000 Image Coding System*, March,1997, http://www.jpeg.org/public/wg1n505.pdf.
[9] Gersho, A. and Gray, R. M., *Vector Quantization and Signal Compression,* Kluwer Academic Publishers, 1991.
[10] Nelson, M., *The Data Compression Book,* 2nd edition., M&T books,November.1995, http://www1.fatbrain.com/asp/bookinfo/bookinfo.asp?theisbn=1558514341
[11] Robert L. Cook and Tony DeRose, *Wavelet Noise*, ACM Transactions on Graphics, July 2005, Volume 24 Number 3, Proc. Of ACM SIGGRAPH 2005, pp. 803-811.
[12] G. Beylkin, R. Coifman, and V. Rokhlin, *Fast wavelet transforms and numerical algorithms*, I. Communications on Pure and Applied Mathematics, 44(2): 141-183, March 1991.
[13] Colm Mulcahy, *Image compression using the Haar Wavelet transforms*, Internal Report.
[14] H. Marmolin, *Subjective MSE measures*, IEEE Trans. Systems Man. Cybernet. 16, 1986, 486–489.
[15] James D. Foley, Andries van Dam, *et. al.Computer Graphics : Principles and Practice*, Addison-Welsey, Reading, MA, 2nd Edition, 1990.